\newcommand{\version}{October 3, 2000}

\documentclass[12pt]{article}
\usepackage{a4,amsthm,amsfonts,latexsym,amssymb}

{\catcode `\@=11 \global\let\AddToReset=\@addtoreset}
\AddToReset{equation}{section}

\swapnumbers
\theoremstyle{plain}
\newtheorem{thm}{THEOREM}[section]
\newtheorem{cl}[thm]{COROLLARY}
\newtheorem{lem}[thm]{LEMMA}

\theoremstyle{definition}
\newtheorem{rem}[thm]{Remark}

\newcommand{\beq}{\begin{equation}}
\newcommand{\eeq}{\end{equation}}
\def\beqa{\begin{eqnarray}}
\def\eeqa{\end{eqnarray}}

\newcommand{\R}{{\mathbb R}}
\newcommand{\C}{{\mathbb C}}
\newcommand{\N}{{\mathbb N}}

\newcommand{\Ll}{{\mathcal L}}
\newcommand{\Hh}{{\mathcal H}}
\newcommand{\Cc}{{\mathcal C}}

\newcommand{\A}{{\bf A}}

\newcommand{\x}{{\bf x}}

\newcommand{\0}{{\bf 0}}

\newcommand{\xpp}{\x_\perp}

\newcommand{\Tr}{{\rm Tr}}
\newcommand{\half}{\mbox{$\frac{1}{2}$}}

\newcommand{\rddm}{\rho^{\rm DDM}}
\newcommand{\rdm}{\rho^{\rm DM}}
\newcommand{\gddm}{\Gamma^{\rm DDM}}
\newcommand{\hddm}{h^{\rm DDM}}
\newcommand{\Eddm}{E^{\rm DDM}}
\newcommand{\Pddm}{\Phi^{\rm DDM}}

\newcommand{\ED}{{\mathcal E}^{\rm DDM}_{B,Z}}
\newcommand{\Elin}{{\mathcal E}^{\rm DDM}_{\rm lin}}
\newcommand{\Econf}{E_{\rm conf}}

\newcommand{\rhs}{\rho^{\rm HS}}
\newcommand{\bsigma}{\mathord{\hbox{\boldmath $\sigma$}}}
\newcommand{\rmd}{{\rm d}}

\date{\small\version}

\begin{document}
\markboth{\scriptsize{HS \version}}{\scriptsize{HS \version}}
\title{\bf{A discrete density matrix theory for atoms in strong magnetic fields}}
\author{\vspace{5pt} Christian Hainzl$^1$ and Robert Seiringer$^2$\\
\vspace{-4pt}\small{Institut f\"ur Theoretische Physik, Universit\"at Wien}\\
\small{Boltzmanngasse 5, A-1090 Vienna, Austria}}

\maketitle

\begin{abstract}
This paper concerns the asymptotic ground state properties of
heavy atoms in strong, homogeneous magnetic fields. In the limit
when the nuclear charge $Z$ tends to $\infty$ with the magnetic
field $B$ satisfying $B \gg Z^{4/3}$ all the electrons are
confined to the lowest Landau band. We consider here an
energy functional, whose variable is a
sequence of one-dimensional density matrices corresponding to
different angular momentum functions in the lowest Landau band. We
study this functional in detail and derive various interesting
properties, which are compared with the density matrix (DM) theory 
introduced by Lieb, Solovej and Yngvason. In contrast to the DM theory the
variable perpendicular to the field is replaced by the discrete angular
momentum quantum numbers. Hence we call the new functional a {\it discrete
density matrix (DDM) functional}. We relate this DDM
theory to the lowest Landau band quantum mechanics and show that it
reproduces correctly the ground state energy apart from errors due
to the indirect part of the Coulomb interaction energy.
\end{abstract}

\footnotetext[1]{E-Mail: \texttt{hainzl@doppler.thp.univie.ac.at}}
\footnotetext[2]{E-Mail: \texttt{rseiring@ap.univie.ac.at}}

\section{Introduction}

The ground state properties of atoms in strong magnetic fields
have been the subject of intensive mathematical studies
during the last decade. This paper is based on the
comprehensive work of Lieb, Solovej and Yngvason \cite{LSY94a,LSY94b}, which
we refer to for an
extensive list of references concerning the history  of this subject.

The starting point of our investigation is the Pauli Hamiltonian
for an atom with $N$ electrons and nuclear charge $Z$ in a
homogeneous magnetic field ${\bf {B}} = (0,0,B)$ with vector
potential ${\bf A}(\x) = \half {\bf B}\times\x$,
\begin{equation}
\label{pho} H  =  \sum_{1 \leq j \leq N} \left\{
\left[\bsigma_j\cdot\left(-i\nabla_j + {\bf
A}(\x_j)\right)\right]^{2} - \frac{Z}{|\x_j|} \right\} +  \sum_{1
\leq i<j \leq N} \frac{1}{|\x_i - \x_j|},
\end{equation}
which acts on the Hilbert space $\bigwedge_{1 \leq j \leq N}
\Ll^{2}({\mathbb {R}}^{3};\C^2)$ of antisymmetric spinor-valued wave functions.
Here $\bsigma$ denotes the usual Pauli spin matrices.

The units are chosen such that $\hbar=2m_{e}=e=1$, so the unit of
energy is four times the Rydberg energy. The magnetic field $B$ is
measured in units $B_0 = \frac{m_{e}^{2}e^{3}c}{\hbar^{3}} = 2.35
\cdot 10^{9} $ Gau\ss, the field strength for which the cyclotron
radius $\ell_{B} = (\hbar c/(eB))^{1/2}$ equals the Bohr radius
$a_{0} = \hbar^{2}/(m_{e}e^{2})$.

The ground state energy is defined as
\beq
\label{gse} E^{\rm Q}(N,Z,B) = \inf\{ \langle\Psi,H\Psi\rangle:
\Psi \in {\rm domain} \,\, H, \langle\Psi,\Psi\rangle = 1\},
\eeq
and if there is a ground state wave function $\Psi$, the corresponding
ground state density $\rho^{\rm Q}$ is given by
\beq
\rho^{\rm Q}(\x) = \rho_\Psi(\x) \equiv N\sum_{s_i=\pm 1/2}\int
|\Psi(\x,s_1,\dots,\x_N,s_N)|^{2} \rmd\x_2\dots\rmd\x_N.
\eeq

Recall that the spectrum of the free Pauli Hamiltonian on
$\Ll^{2}(\R^3;\C^2)$ for one electron in the magnetic field ${\bf B}$,
\beq
H_\A =  \left[\bsigma\cdot(-i\nabla + {\bf A}(\x))\right]^{2},
\eeq
is given by
\beq
p_z^2 + 2\nu B  \qquad \nu = 0,1,2,...,\quad p_z\in\R.
\eeq
The projector $\Pi_0$ onto the lowest Landau band, $\nu = 0$, is
represented by the kernel
\begin{equation}
\label{pk} \Pi_{0}(\x,\x') =
\frac{B}{2\pi}\exp\left\{\frac{i}{2}(\xpp \times \xpp') \cdot {\bf
B} - \frac{1}{4}(\xpp - \xpp')^{2}B\right\}\delta(z -
z')P_\downarrow,
\end{equation}
where $\xpp$ and $z$ are the components of $\x$ perpendicular and
parallel to the magnetic field, and $P_\downarrow$ denotes the
projection onto the spin-down ($s=-1/2$) component. With the
decomposition $\Ll^2(\R^3,\rmd\x;\C^2)=\Ll^2(\R^2,\rmd\xpp)\otimes
\Ll^2(\R,\rmd z)\otimes \C^2$ it can be written as
\beq
\Pi_0=\sum_{m\geq 0}|\phi_m\rangle\langle\phi_m|\otimes{I}\otimes P_\downarrow,
\eeq
where $\phi_m$ denotes the function in the
lowest Landau band with angular momentum $-m\leq 0$, i.e., using polar
coordinates $(r,\varphi)$,
\beq\label{amf}
\phi_m(\xpp)=\sqrt\frac B{2\pi}\frac 1{\sqrt{m !}}\left(\frac
{Br^2}{2}\right)^{m/2}e^{-i m \varphi}e^{-B r^2/4}.
\eeq

The projector onto the subspace of $\bigwedge_{1 \leq j \leq N}
\Ll^{2}({\mathbb {R}}^{3};\C^2)$ where all the electrons are in
the lowest Landau band is the $N$-th tensorial power of $\Pi_0$
and will be denoted by $\Pi_0^N$. The ground state energy of
electrons restricted to the lowest Landau band is defined as
\beq\label{ce}
E^{\rm Q}_{\rm conf}(N,Z,B) = \inf_{\parallel \Psi
\parallel = 1}\langle\Psi, \Pi^{N}_{0} H\Pi^{N}_{0} \Psi\rangle.
\eeq
The assertion that for $B \gg Z^{4/3}$ the electrons are to the
leading order confined to the lowest Landau band is confirmed by
the following theorem.

\begin{thm}[Lowest Landau band confinement]
(\cite{LSY94a}, Theorem 1.2) For any fixed $\lambda=N/Z$ there is
a $\delta(x)$ with $\delta(x)\to 0$ as $x\to\infty$ such that
\beq
\Econf^{\rm Q}\geq E^{\rm Q}\geq \Econf^{\rm
Q}\left(1+\delta(B/Z^{4/3})\right).
\eeq
\end{thm}

We now define the functional we are considering. We are only
interested in $B\gg Z^{4/3}$, so we can restrict ourselves to
considering all particles in the lowest Landau band. Recalling
that
\beq
\Pi_{0}H_{\A}\Pi_{0} = \sum_{m\geq 0}|\phi_m\rangle\langle\phi_m |
\otimes (-\partial_z^2)\otimes P_\downarrow,
\eeq
it is natural to define the following  {\it discrete density
matrix functional} (DDM) as
\beq\label{ddmf}
\ED[\Gamma]=\sum_{m\in\N_0}\left(\Tr[-\partial_z^2 \Gamma_m]-Z\int
V_m(z)\rho_m(z)\rmd z\right)+\widetilde D(\rho,\rho),
\eeq
where
\beq
\widetilde D(\rho,\rho)= \frac 12 \sum_{m, n}\int
V_{m,n}(z-z')\rho_m(z)\rho_n(z')\rmd z \rmd z',
\eeq
and the potentials $V_m$ and $V_{m,n}$ are given by
\beqa\label{defvn}
V_m(z)&=& \int \frac 1{|\x|}|\phi_m(\xpp)|^2 \rmd\xpp, \\ \label{defvnm}
V_{m,n}(z-z') &=& \int \frac {|\phi_m(\xpp)|^2 |\phi_n(\xpp
')|^2}{|\x-\x'|} \rmd\xpp \rmd\xpp'.
\eeqa
Here $\Gamma$ is a sequence of density matrices acting on
$\Ll^2(\R,\rmd z)$,
\beq
\Gamma=(\Gamma_m)_{m\in\N_0},
\eeq
with corresponding densities $\rho=(\rho_m)_m$,
$\rho_m(z)=\Gamma_m(z,z)$. Note that $\ED$ depends on $B$ via the
potentials $V_m$ and $V_{m,n}$.

This functional is defined for all $\Gamma$ with the properties:
\begin{enumerate}
\item[(i)]
\beq
0\leq \Gamma_m \leq I \qquad {\rm for\ all\ }m\in\N_0,
\eeq
\item[(ii)]
\beq
\sum_m\Tr[(1-\partial_z^2)\Gamma_m]<\infty.
\eeq
\end{enumerate}
The corresponding energy is given by
\beq\label{ddme}
E^{\rm DDM}(N,Z,B)=\inf\left\{\ED[\Gamma]\left| \
\sum_m\Tr[\Gamma_m]\leq N\right.\right\}.
\eeq
(We do not require that $N$ is an integer.)
As we will show, $\Eddm$ correctly reproduces the confined ground state
energy $\Econf^{\rm Q}$ apart from errors due to the indirect part
of the Coulomb interaction energy:

\begin{thm}[Relation of $\Eddm$ and $\Econf^{\rm Q}$]\label{11}
For some constant $c_\lambda$ depending only on $\lambda=N/Z$
\beq
0\geq  E_{\rm conf}^{\rm Q}(N,Z,B)-\Eddm(N,Z,B) \geq -R_L,
\eeq
with
\beq\label{rl}
R_L = c_\lambda\min\left\{Z^{17/15}B^{2/5},
Z^{8/3}(1+[\ln(B/Z^{3})]^{2})\right\}.
\eeq
\end{thm}
The proof will be given in Sections \ref{upp} and \ref{lower}.
Note that $\Econf^{\rm Q}$ is of order $\min\{Z^{7/3}
(B/Z^{4/3})^{2/5},Z^3 (1+[\ln(B/Z^3)]^2)\}$, so $R_L$ is really of
lower order.

The functional (\ref{ddmf}) is in fact a {\it reduced
Hartree-Fock} approximation (in the sense of \cite{S91})
for the quantum mechanical many body
problem, i.e. Hartree-Fock theory with the exchange term dropped.
The upper bound in Theorem \ref{11} holds quite generally 
for (reduced) Hartree-Fock
approximations to many body quantum mechanics
with positive two-body interactions.

Hartree and Hartree-Fock approximations to $H$ with restriction to the
lowest Landau level
were probably for the first time considered in \cite{CLR70} and  
\cite{CR73}. They have been studied numerically in
\cite{NLK86,NKL87}.

In this paper we present a rigorous mathematical treatment of the
functional (\ref{ddmf}). In Section \ref{prop} we show that there
exists a unique solution to the minimization problem in
(\ref{ddme}). The corresponding minimizer is composed of
eigenfunctions of one-dimensional effective mean-field
Hamiltonians depending on $m$. Moreover, the superharmonicity of
the effective mean-field potential implies monotonicity and
concavity of the eigenvalues in $m$, which amounts to \lq\lq
filling the lowest angular momentum channels\rq\rq. This fact is important for
numerical treatments of the model, for it means that at most the $N$ lowest
angular momenta have to be considered. For $B/Z^3$
large enough, we will show that each angular momentum channel is
occupied by at most {\it one} particle. In Section \ref{mi} we
estimate the maximum number of electrons that can be bound to the
nucleus. We use Lieb's strategy to derive an upper bound analogous
to \cite{S00}.

The DDM theory
can also be considered as a discrete analogue of
the DM functional introduced in \cite{LSY94a}. To express this
analogy we will recall its definition and main properties in the
next section.

\section{Comparison with the DM functional}\label{compar}

In \cite{LSY94a} Lieb, Solovej and Yngvason defined a density
matrix (DM) functional as
\beq
{\mathcal {E}}^{\rm DM}[\Gamma] = \int_{\R^2}\Tr_{\Ll^2(\R)}
[-\partial_{z}^{2}\Gamma_{\xpp}]\rmd\xpp - Z\int
|\x|^{-1}\rho_{\Gamma}(\x) + D(\rho_{\Gamma},\rho_{\Gamma}).
\eeq
Its variable is an operator valued function
\beq
\Gamma: \xpp \rightarrow \Gamma_{\xpp},
\eeq
where $\Gamma_{\xpp}$ is a density matrix on $\Ll^{2}(\R)$, given
by a kernel $\Gamma_{\xpp}(z,z')$ and satisfying
\beq\label {opb}
 0 \leq \Gamma_{\xpp} \leq (B/2\pi)I
\eeq
as an operator on $\Ll^2(\R)$. Here we denote
\beq
\rho_\Gamma(\x) = \Gamma_{\xpp}(z,z) \quad  {\rm and}\quad D(f,g)
=\frac{1}{2}\int\frac{f(\x)g(\x')}{|\x-\x'|}\rmd\x \rmd\x'.
\eeq
The energy
\beq
E^{\rm DM}(N,Z,B) = \inf\{{\mathcal {E}}^{\rm DM}[\Gamma]|
\,\Gamma \,\mbox{satisfies} \,(\ref {opb})\,\,{\rm and}\,\, \int
\rho_{\Gamma} \leq N\}
\eeq
turns out to be asymptotically equal to the confined quantum
mechanical ground state energy $\Econf^{\rm Q}$ in the following
precise sense:

\begin{thm}[Relation of $E^{\rm DM}$ and $\Econf^{\rm Q}$]\label{21}
(\cite{LSY94a}, Sect. 5 and 8) For some constants $c_\lambda$ and
$c_\lambda'$, depending only on $\lambda=N/Z$,
\beq
R_U \geq E_{\rm conf}^{\rm Q}(N,Z,B)-E^{\rm DM}(N,Z,B) \geq -R_L,
\eeq
with $R_L$ given in (\ref{rl}) and
\beq
R_U=c_\lambda'\min\{Z^{5/3}B^{1/3},
Z^{8/3}[1+\ln(Z)+(\ln(B/Z^{3}))^{2}]^{5/6}\}.
\eeq
\end{thm}

The DM energy fulfills the simple scaling relation
\beq\label{scaldm}
E^{\rm DM}(N,Z,B)=Z^3 E^{\rm DM}(\lambda,1,\eta),
\eeq
where we introduced the parameters $\lambda=N/Z$ and
$\eta=B/Z^3$. In the limit $\eta\to\infty$, $E^{\rm
DM}/(\ln\eta)^2$ converges to the so-called  {\it hyper-strong}
(HS) energy $E^{\rm HS}(\lambda)$, which is the ground state
energy of the functional
\beq\label{hs}
{\cal E}^{\rm HS}[\rho ]=\int_\R \left(\frac{\rmd}{\rmd z}
\sqrt{\rho (z)}\right)^2 \rmd z- \rho (0) + \frac{1}{2}\int \rho
(z)^2\rmd z
\eeq
under the condition $\int \rho(z)\rmd z\leq \lambda$. The
corresponding minimizer can be given explicitly, namely
\begin{eqnarray}\nonumber
\rho^{\rm HS}(z) &=& \frac{2(2-\lambda)^2} {(4\sinh [(2-\lambda
)|z|/4+c(\lambda)])^2} \qquad {\rm for} \quad \lambda <2, \\
\nonumber && \qquad \tanh c(\lambda) = (2-\lambda )/2,
\\ \label{RHS} \rho^{\rm HS}(z) &=& 2(2+|z|)^{-2} \qquad \qquad \qquad
\qquad\quad\ \ {\rm for} \quad \lambda \geq 2.
\end{eqnarray}

We now discuss the relation of $\ED$ and ${\cal E}^{\rm DM}$. The
functional $\ED$ is the restriction of ${\cal E}^{\rm DM}$ to
density matrices of the form
\beq
\Gamma_{\xpp}(z,z')=\sum_m |\phi_m(\xpp)|^2 \Gamma_m(z,z').
\eeq
Therefore it is clear that
\beq\label{ddmgdm}
\Eddm(N,Z,B)\geq E^{\rm DM}(N,Z,B).
\eeq
In the limit
$N\to\infty$ they coincide, i.e. we will show that
\beq\label{limdm}
\lim_{N\to\infty} \frac{\Eddm(N,Z,B)}{Z^3}=E^{\rm
DM}(\lambda,1,\eta)
\eeq
for all fixed $\lambda$ and $\eta$. In fact, (\ref{limdm}) follows immediately
from Theorem \ref{11} and Theorem \ref{21}, but it could of course also be shown
directly without referring to the relation to $\Econf^{\rm Q}$.  
Note that in contrast to DM, the DDM
energy depends non-trivially on all the three parameters $N$, $Z$ and $B$, like
the original QM problem.

\section{Properties of $\ED$}\label{prop}

\begin{thm}[Existence of a minimizer]
For each $N>0$, $B>0$ and $Z>0$ there exists a minimizer $\gddm$
for $\ED$ under the condition $\sum_m \Tr[\Gamma_m]\leq N$.
\end{thm}

\begin{proof}
Let $\Gamma^{(i)}$ be a minimizing sequence for $\ED$ under the
normalization condition $\sum_m \Tr[\Gamma_m]\leq N$, with
corresponding densities $\rho^{(i)}=(\rho_m^{(i)})_m$. Using the
one-dimensional Lieb-Thirring  inequality \cite{LT76}  we have
\beq\label{3norm}
\int\left|\rho_m^{(i)}\right|^3\leq {\rm const.}
\Tr[-\partial_z^2\Gamma_m^{(i)}].
\eeq
Moreover (cf. \cite{LSY94a}, Eq. (4.5b)),
\beq\label{sqrt}
\int\left(\frac{\rmd\sqrt{\rho_m^{(i)}(z)}}{\rmd z}\right)^2\rmd z
\leq \Tr[-\partial_z^2 \Gamma_m^{(i)}].
\eeq
Since the potential energy is relatively bounded with respect to
the kinetic energy (compare Thm. 2.2 in \cite{LSY94a}), the right
hand sides of (\ref{3norm}) and (\ref{sqrt}) are uniformly bounded.
Hence the sequence $\rho_m^{(i)}$ is bounded in $\Ll^3(\R,\rmd
z)\cap \Ll^1(\R,\rmd z)$, and $\sqrt{\rho_m^{(i)}}$ is bounded in
$\Hh^1(\R,\rmd z)$. By the \lq\lq diagonal sequence trick\rq\rq,
there is a subsequence, again denoted by $\rho^{(i)}$, such that
$\rho_m^{(i)}$ converges to some $\rho_m^{(\infty)}$ for each $m$,
weakly in $\Ll^3(\R,\rmd z)\cap \Ll^p(\R,\rmd z)$ for some
$1<p\leq 3$ and pointwise almost everywhere. By Fatou's lemma
\beqa\nonumber &&\liminf_{i\to\infty} \sum_{m,n} \int \rmd z \rmd z'
V_{m,n}(z-z') \rho_m^{(i)}(z) \rho_n^{(i)}(z)\\ &&\geq \sum_{m,n}
\int \rmd z \rmd z' V_{m,n}(z-z') \rho_m^{(\infty)}(z)
\rho_n^{(\infty)}(z). \eeqa Observe now that $V_m\leq C$ for some
$C>0$ and for all $m$. Moreover, $V_m(z)\leq 1/|z|$, so $V_m\in\Ll^p$ for all
$p>1$. By the weak
convergence we can conclude that
\beq
\lim_{i\to\infty} \int \rmd z V_m(z)\rho_m^{(i)}(z)=\int \rmd z
V_m(z)\rho_m^{(\infty)}(z)
\eeq
for each $m$. Moreover, by the dominated convergence theorem, we
get
\beq
\lim_{i\to\infty} \sum_m \int \rmd z V_m(z)\rho_m^{(i)}(z)= \sum_m
\int \rmd z V_m(z)\rho_m^{(\infty)}(z).
\eeq
Using the \lq\lq diagonal
sequence trick\rq\rq\ once more, we can use the trace class
property of the $\Gamma_m^{(i)}$'s to conclude that there exists a
subsequence of $\Gamma^{(i)}$ and a $\gddm=(\gddm_m)_m$ such that
\beq
\Gamma_m^{(i)}\rightharpoonup \gddm_m
\eeq
in weak operator sense, for each $m$. It follows from weak
convergence that $0\leq\gddm_m\leq 1$. Using Fatou's lemma twice,
we have \beq \sum_m \Tr[\gddm_m]\leq \liminf_{i\to\infty}
\sum_m\Tr[\Gamma_m^{(i)}]\leq N. \eeq By the same argument \beq
\sum_m \Tr[-\partial_z^2 \gddm_z]\leq \liminf_{i\to\infty} \sum_m
\Tr[-\partial_z^2\Gamma_m^{(i)}]. \eeq Let now $\rddm_m$ denote
the density of $\gddm_m$. It remains to show that
$\rho_m^{(\infty)}=\rddm_m$ for each $m$. From weak convergence it
follows that \beq\label{WEAK}
(1-\partial_z^2)^{1/2}\Gamma_m^{(i)}(1-\partial_z^2)^{1/2}\rightharpoonup
(1-\partial_z^2)^{1/2}\gddm_m(1-\partial_z^2)^{1/2} \eeq weakly on
the dense set $\Cc_0^\infty(\R)$. Since the operators are bounded
by $\Tr[(1-\partial_z^2)\Gamma_m^{(i)}]\leq C$, we see that
(\ref{WEAK}) holds weakly in $\Ll^2(\R,\rmd z)$. With
$\eta\in\Cc_0^\infty(\R)$ considered as a multiplication operator,
it is easy to see that
$(1-\partial_z^2)^{-1/2}\eta(1-\partial_z^2)^{-1/2}$ is a compact
operator (it is even Hilbert-Schmidt). Thus it can be approximated
in norm by finite rank operators. Using (\ref{WEAK}) we can
therefore conclude that \beq \lim_{i\to\infty}
\Tr[\Gamma_m^{(i)}\eta]=\Tr[\gddm_m\eta], \eeq i.e.
$\rho_m^{(i)}\to\rddm_m$ in the sense of distributions. Since we
already know that $\rho_m^{(i)}$ converges to $\rho_m^{(\infty)}$
pointwise almost everywhere, we conclude that
$\rho_m^{(\infty)}=\rddm_m$ for each $m$.

We have thus shown that there exists a $\gddm$ with $\sum_m
\Tr[\gddm_m]\leq N$ and $\ED[\gddm]\leq
\liminf_{i\to\infty}\ED[\Gamma^{(i)}]=E^{\rm DDM}$.
\end{proof}

\begin{lem}[Uniqueness of the density]
\, The density corresponding to\linebreak the minimizer is unique,
i.e., if there are two minimizers $\Gamma^{(1)}$ and
$\Gamma^{(2)}$, their densities $\rho_m^{(1)}$ and $\rho_m^{(2)}$
are equal, for all $m$.
\end{lem}

\begin{proof}
Observe that
\beq
\widetilde D(\rho,\rho)=D( \tilde\rho,\tilde\rho),
\eeq
where we set $\tilde\rho(\x)=\sum_m |\phi_m(\xpp)|^2 \rho_m(z)$.
Using the positive definiteness of the Coulomb kernel and the fact
that $\tilde\rho^{(1)}=\tilde\rho^{(2)}$ implies
$\rho_m^{(1)}=\rho_m^{(2)}$ for all $m$, we immediately get the
desired result.
\end{proof}

Having established the uniqueness of the density, we can now
define a {\it linearized DDM functional} by
\beq
\Elin[\Gamma]=\sum_m \Tr[\hddm_m \Gamma_m],
\eeq
with the one-particle operators
\beq\label{defhm}
\hddm_m=-\partial_z^2-\Pddm_m(z),
\eeq
where the potentials are given by
\beq\label{defphim}
\Pddm_m(z)=ZV_m(z)-\sum_n \int \rddm_n(z')V_{m,n}(z-z')\rmd z'.
\eeq

\begin{lem}[Equivalence of linearized theory]
A minimizer $\gddm$ for $\ED$ under the constraint $\sum_m
\Tr[\Gamma_m]\leq N$ is also a minimizer for the linearized
functional $\Elin$ (with the same constraint).
\end{lem}
\begin{proof}
We proceed essentially as in \cite{LSY94a}. For any $\Gamma$
\beq\label{essent}
\ED[\Gamma]=\Elin[\Gamma]+\widetilde
D(\rho_\Gamma-\rddm,\rho_\Gamma-\rddm)-\widetilde D(\rddm,\rddm).
\eeq
In particular, for all $\delta>0$,
\begin{eqnarray}\nonumber
\ED[(1-\delta)\gddm+\delta\Gamma]&=&(1-\delta)\Elin[\gddm]+\delta\Elin
[\Gamma]-\widetilde D(\rddm,\rddm)\\ &&+\delta^2\widetilde
D(\rho_\Gamma-\rddm,\rho_\Gamma-\rddm).
\end{eqnarray}
Now if there exists a $\Gamma^{(0)}$ with
$\sum_m\Tr[\Gamma^{(0)}_m] \leq N$ and
$\Elin[\Gamma^{(0)}]<\Elin[\gddm]$ we can choose $\delta$ small
enough to conclude that
\beq
\ED[(1-\delta)\gddm+\delta\Gamma^{(0)}]<\Elin [\gddm]-\widetilde
D(\rddm,\rddm)=\ED[\gddm],
\eeq
which contradicts the fact that $\gddm$ minimizes $\ED$.
\end{proof}

We now are able to construct the minimizer of $\ED$ by means of
the eigenfunctions $e^i_m$ of the one-dimensional operators
$\hddm_m$. If $\mu_m^1<\mu_m^2<\dots$ denote the corresponding eigenvalues,
there is a $\mu\leq 0$ such that
\beq\label{gddm}
\gddm_m=\sum_{i=1}^{I_m-1} |e^i_m\rangle\langle e^i_m| + \lambda_m
|e^{I_m}_m\rangle\langle e^{I_m}_m|,
\eeq
where $I_m=\max\{i\, :\, \mu_m^i\leq\mu\}$, and $0\leq
\lambda_m\leq 1$ is the filling of the last level. Since
$\lambda_m$ is determined by the unique density $\rddm_m$, we
immediately get the following corollary:

\begin{cl}[Uniqueness of $\gddm$]
For any $B>0$ and $Z>0$ the minimizer of the functional $\ED$
under the condition $\sum_m \Tr[\Gamma_m]\leq N$ is unique.
\end{cl}

Note that $\gddm$ really depends on the three parameters $N$, $Z$ and
$B$, but we suppress this dependence for the simplicity of the notation.

Of course, the choice of $\mu$ is not unique, but $I_m$ is unique for every $m$.
In the following, we will choose
for $\mu$ the smallest possible, i.e.
\beq
\mu \equiv \max\{\mu_m^{I_m} \}.
\eeq
The energy $E^{\rm DDM}(N,Z,B)$ is a convex, non-increasing
function in $N$. Moreover, it has the following property.

\begin{thm}[Differentiability of $\Eddm$]\label{diffthm}
For $N\not\in \N$ the energy $E^{\rm DDM}$ is differentiable in
$N$, and the derivative is given by $\partial E^{\rm DDM}/\partial
N=\mu$ with $\mu$ given above. For $N\in\N$, the
right and left derivatives, $\partial\Eddm/\partial N_{\pm}$, are
given by
\beq
\frac{\partial\Eddm}{\partial N_-}=\mu,\quad \frac{\partial\Eddm}{\partial
N_+}=\min\{\mu_m^{I_m+1}\}.
\eeq
\end{thm}

\begin{proof}
For fixed $N$, $Z$ and $B$ define $\Elin$ as above. Let $E_{\rm
lin}(N')$ be the infimum of $ \Elin$  under the constraint
$\sum_m\Tr[\Gamma_m]\leq N'$. For $0<N_-<N<N_+$ let $\Gamma^\pm$
be the corresponding minimizers. We have, with $0<\delta<1$,
\beqa\nonumber
\Eddm(N+\delta(N_\pm -N),Z,B)&\leq&
\ED[(1-\delta)\gddm+\delta\Gamma^\pm]\\ \nonumber &=&
\Eddm(N,Z,B)\\&&+\delta\left(E_{\rm lin}(N_\pm)-E_{\rm
lin}(N)\right)+O(\delta^2).
\eeqa
Dividing by $\delta(N_\pm-N)$ and taking the limit $\delta\to 0$ followed
by $N_\pm\to N$,
we conclude that
\beq\label{ineq}
\frac{\partial E_{\rm lin}}{\partial N_-}\leq \frac{\partial \Eddm}
{\partial N_-}, \qquad \frac{\partial \Eddm}
{\partial N_+}\leq \frac{\partial E_{\rm lin}}
{\partial N_+}.
\eeq
{}From (\ref{essent}) we infer that $\Eddm(N',Z,B)\geq E_{\rm
lin}(N')-\widetilde D(\rddm,\rddm)$ for all $N'$, with equality for
$N=N'$. Therefore we get the inverse inequalities in (\ref{ineq}), so
equality holds.
The assertions of the theorem follow
from
\beq
E_{\rm lin}(N)=\sum_{j=1}^{[N]} \mu(j) + \left(N-[N]\right)
\mu([N]+1),
\eeq
where $[N]$ is the largest integer $\leq N$, and $\mu(j)$ is the $j$'th element
of the set $\{\mu_m^i\}$ in increasing order.
\end{proof}

Corresponding to the DDM minimizer we define the {\it
three-dimensional DDM density} as
\beq
\tilde\rddm(\x)=\sum_{m\geq 0}|\phi_m(\xpp)|^2 \rddm_m(z).
\eeq

\begin{thm}[Ordering of the $\mu_m^i$]\label{ordering}
Assume $\hddm_m$ has at least $M$ eigenvalues. Then, for $m\geq
1$,
\beq\label{ordermu}
\sum_{i=1}^M\left(m \mu_{m-1}^i+(m+1)\mu_{m+1}^i
-(2m+1)\mu_m^i\right)\leq -\frac {2\pi}B \sum_{i=1}^M \int
\tilde\rddm |\phi_m e_m^i|^2
\eeq
(with the understanding that $\mu_{m-1}^i$ {\rm
(}$\mu_{m+1}^i${\rm )} $=0$ if $\hddm_{m-1}$ {\rm
(}$\hddm_{m+1}${\rm )} has less than $i$ eigenvalues). The
analogous inequality for $m=0$ is
\beq\label{ordermu2}
\sum_{i=1}^M\left(\mu_{1}^i -\mu_0^i\right)\leq Z \sum_{m=1}^M
|e_0^i(0)|^2 -\frac {2\pi}B \sum_{i=1}^M \int \tilde\rddm |\phi_0
e_0^i|^2.
\eeq
\end{thm}

\begin{proof}
With $\Delta_{(2)}$ the two-dimensional Laplacian, one easily
computes that
\beq\label{327}
(2B)^{-1}\Delta_{(2)} |\phi_m|^2 = m |\phi_{m-1}|^2 + (m+1) |\phi_{m+1}|^2 -
(2m+1)
|\phi_m|^2
\eeq
for $m\geq 1$. Multiplying (\ref{327}) with
$\Phi(\x)=Z/|\x|-\tilde\rddm*|\x|^{-1}$ and integrating over $\xpp$, we
therefore get, for any density matrix $\gamma$ (recall the definitions 
(\ref{defvn}), (\ref{defvnm}), (\ref{defhm}) and (\ref{defphim})),
\beqa\nonumber
&&\Tr\left[\left( m
\hddm_{m-1}+(m+1)\hddm_{m+1}-(2m+1)\hddm_m\right)\gamma\right]\\
\nonumber &&=-(2B)^{-1}\int \rmd\x \Phi(\x) \rho_\gamma(z)
\Delta_{(2)} |\phi_m(\xpp)|^2\\  \label{222} &&=-\frac{2\pi} B
\int \rmd\x \tilde \rddm |\phi_m|^2\rho_\gamma +\frac 1{2B}\int
\rmd z \Pddm_m\frac{\partial^2}{\partial z^2} \rho_\gamma,
\eeqa
where we used partial integration for the second step, and the fact that
$\phi_m(\0)=0$ for $m\geq 1$. To treat the last
term in (\ref{222}), note that the function
\beq\label{functw}
w\to \sum_{i=1}^M \int \rmd z \Pddm_m(z) |e_m^i(z+w)|^2
\eeq
has its maximum at $w=0$, because otherwise one could lower the
energy by shifting the eigenvectors $e_m^i$. Therefore the second
derivative of (\ref{functw}) at $w=0$ is negative. Setting
 $\gamma=\sum_{i=1}^M |e_m^i\rangle\langle e_m^i|$ we see that the last
term in (\ref{222}) is negative, so we can conclude that
\beq
\Tr\left[\left( m
\hddm_{m-1}+(m+1)\hddm_{m+1}-(2m+1)\hddm_m\right)\gamma\right]\leq
-\frac{2\pi} B \int \tilde \rddm |\phi_m|^2\rho_\gamma.
\eeq
By the variational principle (\ref{ordermu}) holds. The proof of
(\ref{ordermu2}) is analogous, considering also the contribution
from $\phi_0(\0)$ in (\ref{222}).
\end{proof}

As a corollary, we immediately get

\begin{cl}[Monotonicity of $\sum_{i=1}^M\mu_m^i$ in $m$]\label{monmu}
$\sum_{i=1}^M \mu_m^i$ is increasing in $m$ for all $M$, and
strictly increasing as long as it is $<0$. Moreover, $\gddm_m=0$
for $m\geq N$.
\end{cl}
\begin{proof}
This follows immediately from Theorem \ref{ordering}, noting that
$\lim_{m\to\infty} \mu_m^i = 0$ for all $i$.
\end{proof}

\begin{rem}[Concavity of $\sum_{i=1}^M\mu_m^i$]
The result (\ref{ordermu}) (together with the monotonicity) implies concavity of
$\mu_m^1 $ in $m$. More precisely,
\beqa\nonumber
&&\half \mu_{m-1}^1+\half\mu_{m+1}^1= \\
&&\frac m{2m+1}\mu_{m-1}^1 + \frac{m+1}{2m+1} \mu_{m+1}^1+\frac
1{2(2m+1)} (\mu_{m-1}^1-\mu_{m+1}^1) \leq  \mu_{m}^1
\eeqa
(and the same holds for $\sum_{i=1}^M \mu_m^i$). This is the
analogue of Prop. 2.3 in \cite{LSY94a}, which states that
$-\mu_1(\xpp)$ is a increasing and concave function of $|\xpp|$
(note the different sign convention).
\end{rem}

We now introduce the parameters $\lambda=N/Z$ and $\eta=B/Z^3$. The next
theorem
deals with the $\eta\to\infty$ limit of $E^{\rm DDM}$ with fixed $\lambda$. To
prove it we need the following Lemma.

\begin{lem}[Convergence to delta function]\label{scallem}
Let $L=L(\eta)$ be the solution of the equation
\beq\label{lbeta}
\eta^{1/2}=L(\eta)\sinh(L(\eta)/2).
\eeq
Let $\psi\in \Hh^1(\R,\rmd z)$, with $\lambda=\int|\psi|^2$ and
$T=\int|\rmd\psi/\rmd z|^2$. Then, for all $m\geq 0$, $Z>0$ and
$\eta=B/Z^3$,
\beqa\nonumber
&&\left| |\psi(0)|^2-\frac 1{Z L^2}\int V_m(z/LZ) |\psi(z)|^2 \rmd
z \right| \\ &&\qquad\qquad\qquad\leq \frac 1L \left(\lambda
\sqrt{\frac\pi 2} \frac {Z^{1/2}}{(m+1)^{1/2}}+16 \lambda^{1/4}
T^{3/4} \frac{(m+1)^{1/4}}{Z^{1/4}}\right).
\eeqa
\end{lem}

\begin{proof}
After an appropriate scaling, this
is a direct consequence of \cite{BSY00}, Lemma 2.1, using the estimates
\beq
\int|\phi_m(\xpp)|^2 |\xpp|^{-1}\rmd\xpp\leq \sqrt{\frac\pi
2}\frac {B^{1/2}}{(m+1)^{1/2}}
\eeq
and
\beq
\int|\phi_m(\xpp)|^2 |\xpp|^{1/2}\rmd\xpp\leq 2 \frac
{(m+1)^{1/4}}{B^{1/4}}.
\eeq
\end{proof}

\begin{thm}[The limit $B\gg Z^3$]\label{largeb}
For all $\lambda>0$
\beq\label{ddmhs}
\lim_{\eta\to\infty}\frac{E^{\rm DDM}(\lambda Z,Z,\eta Z^3)}{Z^3
(\ln\eta)^2}=E^{\rm
HS}(\lambda),
\eeq
uniformly in $Z$.
\end{thm}

\noindent{\it Remark:} The uniformity in $Z$ will be important for the proof of
Theorem \ref{rank1}. It is non-trivial in contrast to DM, where one has the
scaling relation (\ref{scaldm}), which implies that the left hand side of
(\ref{ddmhs}) (with DDM replaced by DM) is independent of $Z$.

\begin{proof}
The lower bound is quite easy, using the results of \cite{LSY94a}.
As shown in Section \ref{compar}, we have
\beq\label{ddmdm}
\frac{E^{\rm DDM}(\lambda Z,Z,\eta Z^3)}{Z^3}\geq
\frac{E^{\rm DM}(\lambda Z,Z,\eta Z^3)}{Z^3}=
E^{\rm DM}(\lambda ,1,\eta),
\eeq
where we have used the scaling properties of $E^{\rm DM}$. It is shown in
\cite{LSY94a} that the right hand side of (\ref{ddmdm}) divided by
$(\ln\eta)^2$ converges to $E^{\rm HS}(\lambda)$.

For the upper bound we assume $N\in\N$ for the moment.
We use as trial density matrices
\beqa\nonumber
\Gamma_m(z,z')&=&\frac{Z^2}N L\sqrt{\rhs(L Z z)}\sqrt{\rhs(L Z z')},
\quad 0\leq m
\leq N-1 \\ \label{trial}
\Gamma_m(z,z')&=&0, \quad m\geq N,
\eeqa
where $\rhs$ is given in (\ref{RHS}) and $L=L(\eta)$ is defined in
(\ref{lbeta}). The kinetic energy is easily computed to be
\beq
Z^3 L^2 \int |\rmd\sqrt{\rhs}/\rmd z|^2.
\eeq
For the attraction term we use Lemma \ref{scallem} to estimate
\beqa\nonumber
\frac 1{Z^3 L^2} \sum_m Z\int V_m \rho_m &\geq& \frac 1N \sum_{m=0}^{N-1}\left(
\rhs(0)-\frac {C_\lambda}L\left(\frac {Z^{1/2}}{(m+1)^{1/2}}+ \frac{(m+1)^{1/4}}
{Z^{1/4}}\right)\right) \\
&\geq& \rhs(0)- \frac {C'_\lambda}L
\eeqa
for some constant $C'_\lambda$ depending on $\lambda$.

For the repulsion term we first estimate
\beq
\sum_{0\leq m,n\leq N-1}V_{m,n}(z)\leq \left(\frac
B{2\pi}\right)^2 \int \frac{\Theta(\sqrt{2
N/B}-|\xpp|)\Theta(\sqrt{2
N/B}-|\xpp'|)}{\sqrt{|\xpp-\xpp'|^2+z^2}}\rmd\xpp \rmd\xpp',
\eeq
which follows from monotonicity of $1/|\x|$ in $|\xpp|$ and the
fact that $\sum_m |\phi_m|^2\leq B/2\pi$. Therefore we have
\beqa\nonumber
&&\frac 1{Z^3 L^2} \sum_{0\leq m,n\leq N-1} \int
V_{m,n}(z-z')\rho_m(z)\rho_m(z')\rmd z\rmd z'\\
&&\qquad\qquad\qquad\leq \frac 1L \xi \int f(\xi
(z-z'))\rhs(z)\rhs(z')\rmd z \rmd z',
\eeqa
where we set $\xi=(\eta/\lambda)^{1/2}/L$, and the function $f$
is given by
\beq
f(z)=(2\pi)^{-2}\int \frac{\Theta(\sqrt{2}-|\xpp|)\Theta(\sqrt{2}
-|\xpp'|)} {\sqrt{|\xpp-\xpp'|^2+z^2}}\rmd\xpp \rmd\xpp'.
\eeq
We now claim that $L^{-1}\xi f(\xi z)\to \delta(z)$ as
$\eta\to\infty$. Since $f(z)\leq 1/|z|$ we have, for any
$\chi\in\Hh^1\cap\Ll^1$,
\beqa\nonumber
&&\left| L^{-1}\xi\int f(\xi z)\chi(z)\rmd z-\chi(0)
L^{-1}\int_{-\xi}^\xi f(z) \rmd z \right| \\ \nonumber &&=
\left|L^{-1} \xi \int_{-1}^1 f(\xi z) (\chi(z)-\chi(0))\rmd
z+L^{-1} \xi\int_{|z|\geq 1}f(\xi z)\chi(z)\rmd z\right|\\ && \leq
{\rm const.} L^{-1} \xi \int_{-1}^1 f(\xi z) z^{1/2}\rmd z+L^{-1}
\int_{|z|\geq 1}|\chi(z)|\rmd z\leq {\rm const.} L^{-1}.
\eeqa
Now $L^{-1}\int_{|z|\leq \xi}f \to 1$ as $\eta\to\infty$, which
proves our claim. And since $L(\eta)\approx \ln\eta$ for large
$\eta$, this finishes the proof of Theorem \ref{largeb}, in the case where $N$
is an integer. The proof for $N\not\in \N$ is analogous, using (\ref{trial}) 
with the
density matrix corresponding to $m=[N]$ multiplied by $N-[N]$ as 
trial density matrices.
\end{proof}

\begin{cl}[HS limit of the density]\label{hsdens}
For fixed $\lambda=N/Z$
\beq\label{345}
\lim_{\eta\to\infty} \frac 1 {Z^2 \ln\eta} \sum_m \rddm_m(z/Z\ln\eta)=\rhs(z)
\eeq
in the weak $\Ll^1$ sense, uniformly in $Z$.
\end{cl}

\begin{proof}
The convergence of the densities in (\ref{345}) follows from the
convergence of the energies in a standard way by considering perturbations
of the external
potential (cf. e.g. \cite{LSY94a}). Moreover, since the convergence in
(\ref{ddmhs}) is uniform in $Z$, (\ref{345}) holds for any function
$Z=Z(\beta)$, so we can conclude that (\ref{345}) holds uniformly in $Z$,
too. 
\end{proof}

Using the results above we can now prove the analogue of Theorem 4.6 in
\cite{LSY94a}.

\begin{thm}[$\gddm_m$ has rank at most $1$ for large $\eta$]\label{rank1}
There exists a constant $C$ such that $\eta\geq C$ implies that $\gddm_m$ has
rank at most $1$ for all $m$.
\end{thm}

\begin{proof}
We first treat the case $\lambda<2$. From Theorems \ref{diffthm} and  
\ref{largeb} and the fact that
$\Eddm(\lambda Z,Z,\eta Z^3)$ is convex in $\lambda$, we get
\beq\label{241}
\lim_{\eta\to\infty} \frac{\mu}{Z^2 (\ln\eta)^2}=\frac{d E^{\rm
HS}(\lambda)}{d\lambda}<0.
\eeq
Suppose that $\mu$ is not the ground state energy of some $\hddm_m$.
Then $\mu\geq
-Z^2/4$, because the second lowest eigenvalue of $\hddm_m$ is equal to the
ground state energy of the three-dimensional operator
\beq\label{3d}
-\Delta-\Pddm_m(|\x|).
\eeq
This follows
because $\Pddm_m(z)$ is reflexion symmetric, so the eigenvector
corresponding to the second lowest eigenvalue, $u_m(z)$, has a node at
$z=0$. Therefore $u_m(|\x|)/|\x|$ is an eigenvector of
(\ref{3d}), and because it does not change sign, it must be a
ground state. Since $\Pddm_m(|\x|)\leq Z/|\x|$, the ground state energy of
(\ref{3d}) is greater than $-Z^2/4$.
So $\mu/Z^2(\ln\eta)^2$ would go to zero as
$\eta\to\infty$, in contradiction to (\ref{241}). Therefore there exists a
constant $C$ such that $\eta>C$ implies the assertion to the theorem. This
constant can be chosen independent of $Z$, because the limit
(\ref{241}) is uniform in $Z$ (by the same argument as in the proof of
Corollary \ref{hsdens}).

Now assume that $\lambda=1+\bar\delta$ for some $\bar\delta>0$. From Corollary
\ref{hsdens} we infer that for large enough $\eta$ there is some $c_\delta>0$ such
that
\beq\label{cdelta}
\sum_m \int_{|z|\leq c_\delta (Z\ln\eta)^{-1}} \rddm_m \geq (1+\half\delta)Z,
\eeq
where $\delta=\min\{1,\bar\delta\}$. We now will show that for
$\eta$ large enough $\hddm_m$ has at most one eigenvalue, for all
$0\leq m \leq N$. By the same argument as above, we need to show
that the three-dimensional operator (\ref{3d})
has no eigenvalues. Using $V_m(z)\leq 1/|z|$ and (cf. the next section)
\beq
V_{m,n}(z)\geq \frac 1 {\sqrt 2} V_{m+n}(z/\sqrt 2)\geq \frac
1{\sqrt{z^2+ 2(m+n+1)B^{-1}}} \geq \frac 1{\sqrt{z^2+
2(2N+1)B^{-1}}}
\eeq
we can use (\ref{cdelta}) to estimate
\beq
\Pddm_m(|\x|)\leq \frac Z {|\x|} - \frac{ Z (1+\half\delta) }
{\sqrt{\left(|\x|+c_\delta (Z\ln\eta)^{-1} \right)^2+
2(2N+1)B^{-1}}} .
\eeq
Therefore, for $\eta$ large enough,
\beq
\Pddm_m(|\x|)\leq \left\{\begin{array}{cl} Z/|\x|\quad & {\rm
for}\quad |\x|\leq 3 \delta^{-1} c_\delta (Z\ln\eta)^{-1} \\
0\quad       & {\rm otherwise}. \\
\end{array}\right.
\eeq
By means of the Cwikel-Lieb-Rosenbljum bound \cite{RS78} we can estimate the number
of negative eigenvalues of (\ref{3d}) as
\beq
{\rm const.} \int|\Pddm_m(|\x|)|_+^{3/2}\rmd\x\leq c_\delta'
(\ln\eta)^{-3/2},
\eeq
which is less than 1 for $\eta$ large enough.
\end{proof}

\begin{rem}[Chemical potential for large $B/Z^3$]
The theorem above,\linebreak together with Corollary \ref{monmu}, shows
that for
$B/Z^3$ large
enough, the chemical potential is given by the ground state energy
of $\hddm_{<N-1>}$, i.e. $\mu=\mu_{<N-1>}^1$, where
$<N>$ denotes the smallest integer $\geq N$.
\end{rem}

\section{Maximal negative ionization}\label{mi}

The DDM energy is convex and monotonously decreasing in $N$. Because
$\widetilde D(\rho,\rho)$ is strictly convex in $\rho$, $\Eddm$ is
strictly convex up to some $N=N_c(Z,B)$, and constant for $N\geq
N_c$. By uniqueness of $\gddm$ the minimizer for $N>N_c$ is equal
to the one with $N=N_c$. In particular,
$\sum_m\Tr[\gddm_m]=\min\{N,N_c\}$.

The \lq\lq critical\rq\rq\ $N_c$ measures the maximal particle
number that can be bound to the nucleus. We will proceed
essentially as in \cite{S00} and use Lieb's strategy \cite{L84} to
get an upper bound on $N_c$. In addition, the following Lemma is
needed. Throughout, we use various properties of $V_m$ stated in
\cite{BRW99}, Sect. 4., and proven in \cite{RW00}.

\begin{lem}[Comparison of $V_m$ and $V_{m,n}$]\label{vmvmn}
\beq\label{assert}
\left(\frac 1{V_m(z)}+\frac 1{V_n(z)}+\frac 1{V_m(z')}+\frac 1{V_n(z')}\right)
V_{m,n}(z-z')\geq 1
\eeq
\end{lem}

\begin{proof}
Using the definition of $V_{m,n}$ it can be shown (cf. \cite{P82}) that
\beq
\sqrt 2 V_{m,n}(\sqrt 2 z)=\sum_{i=0}^{m+n} c_i V_i(z)
\eeq
for some coefficients $c_i\geq 0$ that fulfill $\sum_i c_i=1$. 
Since $V_m\geq V_{m+1}$
for all $m$ (\cite{BRW99}, 4b) we get
\beq
V_{m,n}(z)\geq \frac 1{\sqrt 2} V_{m+n}(z/\sqrt 2)\geq \frac 12 V_{m+n}(z/2),
\eeq
where we used the fact that $a V_m(az)\leq V_m(z)$ if $a\leq 1$ 
(\cite{BRW99}, 4g).
Moreover, using convexity of $1/V_{m+n}$ (\cite{BRW99}, 4i), we arrive at
\beq
V_{m,n}(z-z')\geq \left(V_{m+n}(z)^{-1}+V_{m+n}(z')^{-1}\right)^{-1}.
\eeq
The assertion (\ref{assert}) follows if we can show that
\beq
\frac 1 {V_{m+n}(z)}\leq \frac 1 {V_{m}(z)}+\frac 1 {V_{n}(z)}.
\eeq
This is of course trivial if $n$ or $m$ equals zero. If $n,m\geq 1$ we use
$\sqrt{z^2+m}^{-1}\geq V_m(z)\geq \sqrt{z^2+m+1}^{-1}$ (\cite{BRW99}, 4a) to estimate
\beqa \nonumber
\frac 1 {V_{m+n}(z)}&\leq& \sqrt{z^2+m+n+1}\leq 2\sqrt{z^2+(\sqrt m+\sqrt n)^2/4} \\
&\leq& \sqrt{z^2+m}+\sqrt{z^2+n}\leq \frac 1{V_m(z)}+\frac 1{V_n(z)},
\eeqa
which finishes the proof.
\end{proof}

\begin{thm}[Critical particle number]
\beq\label{NC}
Z\leq N_c \leq 4Z-\frac 1{N_c}\frac {\partial \Eddm(N_c,Z,B)}{\partial Z}
\eeq
\end{thm}

\noindent {\it Remark:} The factor $4$ stems from the
symmetrization of (\ref{assert}) in $m$ and $n$. Due to this
symmetrization one could expect that Lemma \ref{vmvmn} holds with
1 replaced by 2 on the right hand side. This would imply that $4Z$
could be replaced by $2Z$ in (\ref{NC}).

\begin{proof}
Let $e_m^i$ denote the eigenvectors of $\hddm_m$, i.e.
\beq\label{heme}
\hddm_m e_m^i = \mu_m^i e_m^i.
\eeq
Multiplying (\ref{heme}) with $e_m^i /V_m$ and integrating we get
\beq\label{48}
Z\geq \sum_n\int \frac 1 {V_m(z)} e_m^i(z)^2
V_{m,n}(z-z')\rddm_n(z')\rmd z\rmd z' +\langle e_m^i| \frac
1{V_m}(-\partial_z^2)| e_m^i\rangle,
\eeq
where we used that $\mu_m^i\leq 0$. Since $1/V_m$ is convex and
$|z|V_m(z)\to 1$ as $|z|\to\infty$ we have $|(1/V_m)'|\leq 1$.
Using this and partial integration we can estimate the last term
in (\ref{48}) by
\beqa\nonumber
\langle e_m^i|\frac 1{V_m}(-\partial_z^2)|e_m^i\rangle &=& \langle
e_m^i V_m^{-1/2}|-\partial_z^2-\frac 14
\left|\frac{V_m'}{V_m}\right|^2
 |e_m^i V_m^{-1/2}\rangle \\
 &\geq& -\frac 14\int V_m(z) e_m^i(z)^2 \rmd z .
\eeqa
Summing over all $m$ and $i$ (and, according to Eq. (\ref{gddm}),
multiplying the factor corresponding to the largest $\mu_m^i$ by
$\lambda_m$), we arrive at
\beq\label{49}
NZ\geq \sum_{m,n}\int \frac 1 {V_m(z)} \rddm_m(z)
V_{m,n}(z-z')\rddm_n(z')\rmd z\rmd z' -\frac 14 \sum_{m}\int V_m
\rddm_m.
\eeq
Note that the last term in (\ref{49}) is equal to
$\partial\Eddm/\partial Z$.  To treat the first term in (\ref{49})
we use symmetry and Lemma \ref{vmvmn} to get
\beqa\nonumber
&&\sum_{m,n}\int \frac 1{V_m(z)}\rddm_m(z) V_{m,n}(z-z')
\rddm_n(z') \rmd z\rmd z' \\ \nonumber &&=\frac 14 \sum_{m,n}\int
\left(\frac 1{V_m(z)}+\frac 1{V_n(z)}+\frac 1{V_m(z')} +\frac
1{V_n(z')}\right)\\ &&\qquad \qquad \times
\rddm_m(z)V_{m,n}(z-z')\rddm_n(z') \rmd z\rmd z' \geq \frac 14
N^2.
\eeqa
Inserting this into (\ref{49}) and dividing by $N/4$ we arrive at
\beq
N\leq 4Z-\frac 1{N}\frac {\partial \Eddm}{\partial Z}.
\eeq

The lower bound on $N_c$ is quite easy. We just have to show that
$\hddm_{[N+1]}$ has a bound state if $N<Z$. Using $\psi(z)=\exp(-a
|z|)$ with $a>0$ as a trial vector we compute
\beqa\nonumber
&&\langle\psi|\hddm_{[N+1]}|\psi\rangle\\ \nonumber &&= a-Z\int
V_{[N+1]}(z)e^{-2a|z|}\rmd z+\sum_n \int
V_{n,[N+1]}(z-z')\rddm_n(z')e^{-2a|z|}\rmd z\rmd z'\\ &&\leq a
-Z\int V_{[N+1]}(z)e^{-2a|z|}\rmd z+N \max_{n\leq N}\int
V_{n,[N+1]}(z)e^{-2a |z|}\rmd z.
\eeqa
Since
\beq
\lim_{a\to 0} \frac 1 {\ln(1/a)} \int V_{[N+1]}e^{-2a|z|}=\lim_{a\to
0} \frac 1{\ln(1/a)} \int V_{n,[N+1]} e^{-2a|z|}=1,
\eeq
$\langle\psi|\hddm_{[N+1]}|\psi\rangle$ will be negative for small
enough $a$, if $N<Z$.
\end{proof}

\begin{rem}[Explicit bound on $N_c$] The concavity of $\Eddm$
in $Z$ implies that $\partial \Eddm(N,Z,B)/\partial Z\geq
\Eddm(N,2Z,B)/Z$. Using (\ref{ddmgdm}) and the bounds on $E^{\rm
DM}$ given in \cite{LSY94a}, Thm. 4.8, (\ref{NC}) implies the
upper bound
\beq\label{uppnc}
N_c\leq 4 Z \left(1+ C \min\left\{(B/Z^3)^{2/5}, 1+[\ln(B/Z^3)]^2\right\}\right)
\eeq
for some constant $C$ independent of $Z$ and $B$.
\end{rem}

\begin{rem}[Upper bound on $N_c^{\rm DM}$]
The upper bound (\ref{uppnc}) holds also for $N_c^{\rm DM}$, the critical
particle number in the DM theory. In fact, the convergence in
(\ref{limdm}) implies that
\beq
N_c^{\rm DM}\leq Z \left(\liminf_{Z\to\infty}
\frac{N_c(Z,\eta Z^3)}{Z}\right)
\eeq
for all fixed $\eta=B/Z^3$.
\end{rem}

\section{Upper bound to the QM energy}\label{upp}

We now show that $\Eddm$ is an upper bound to the quantum
mechanical ground state energy. In fact it is even an upper bound
to $\Econf^{\rm Q}$. By Lieb's variational principle \cite{L81}
\beq
E^{\rm Q}\leq \Tr[(H_\A-Z |\x|^{-1})\gamma]+\frac12 \int
\frac{\gamma(\x,\x) \gamma(\x',\x') - |\gamma(\x,\x')|^2}{
|\x-\x'|}\rmd\x \rmd\x'
\eeq
for all density matrices $0\leq \gamma\leq 1$ with $\Tr[\gamma]\leq N$.
We choose
\beq
\gamma(\x,\x')=\sum_m \phi_m(\xpp)\overline{\phi_m}(\xpp')
\Gamma_m(z,z').
\eeq
Since $\Pi_0\gamma\Pi_0=\gamma$, we get an upper bound even for
$\Econf^{\rm Q}$. Omitting the negative \lq\lq exchange
term\rq\rq, we compute
\beq
\Econf^{\rm Q}(N,Z,B)\leq \ED[\Gamma],
\eeq
where $\Gamma=(\Gamma_m)_m$. Note that $\Tr[\gamma]=\sum_m \Tr[\Gamma_m]$.
Therefore we immediately conclude that
\beq
\Econf^{\rm Q}(N,Z,B)\leq E^{\rm DDM}(N,Z,B).
\eeq

\section{Lower bound to the QM energy}\label{lower}

To get a lower bound on the QM energy we need to estimate the
two-body interaction
potential in terms of one-body potentials. One way to do this is
to use the Lieb-Oxford
inequality \cite{LO81} together with
 the positive definiteness of the Coulomb kernel:
\beq
\langle \Psi|\sum_{i<j}|\x_i-\x_j|^{-1}\Psi\rangle\geq 2
D(\rho,\rho_\Psi)-D(\rho,\rho)-1.68\int\rho_\Psi^{4/3},
\eeq
where we can choose $\rho=\tilde\rddm$. However, if $\Psi$ is an
(approximate) ground state wave function in the lowest Landau
band, the error term $\int\rho_\Psi^{4/3}$ will in general be greater than
the energy itself. More precisely, if $\gamma_\Psi$ denotes the
one-particle density matrix of $\Psi$, we can estimate
\beqa\nonumber
\int\rho_\Psi^{4/3}&\leq& \left( \int\rho_\Psi^3\right)^{1/6}
\left( \int\rho_\Psi\right)^{5/6}\\ \nonumber &\leq& \left(\frac
3{\pi^2} B^2 \Tr[-\partial_z^2 \gamma_\Psi]\right)^{1/6} N^{5/6}\\
&\leq& {\rm const.} Z^{1/5} N^{14/15} B^{2/5},
\eeqa
where we used \cite{LSY94a}, Lemma 4.2, to estimate
$\int\rho_\Psi^3$ and the kinetic energy. For large enough $B$
this bound is of no use.

That is why we follow the method of \cite{LSY94a} to get an improved bound for
large $B$. Their result is that for wave functions $\Psi$ that satisfy $\langle\Psi
| H \Psi\rangle <0$ the bound
\beq\label{impr}
\langle \Psi|\sum_{i<j}|\x_i-\x_j|^{-1}\Psi\rangle\geq 2
D(\rdm,\rho_\Psi)-D(\rdm,\rdm)-c_\lambda Z^{8/3}\left(1+(\ln\eta)^2\right)
\eeq
holds. The proof of this result uses only properties of $\rdm$ which hold also
for $\tilde\rddm$. Therefore
(\ref{impr}) holds also with $\rdm$ replaced by $\tilde\rddm$,
possibly with a different constant. So we can estimate
\beqa\nonumber
\Econf^{\rm Q}(N,Z,B)&\geq&\inf_\Psi
\langle\Psi|\sum_i\left(H_\A^{(i)}-Z/|\x_i|+\tilde\rddm*|\x_i|^{-1}\right)\Psi\rangle
\\ &&-D(\tilde\rddm,\tilde\rddm)-R_L,
\eeqa
where the infimum is over all $\Psi$ with $\Pi_0^N\Psi=\Psi$ and $\|\Psi\|=1$, and
\beq
R_L=c_\lambda \min\{Z^{17/15} B^{2/5},Z^{8/3}\left(1+(\ln\eta)^2\right)  \}.
\eeq
Because $H_\A-Z/|\x|+\tilde\rddm*|\x|^{-1}$ is a one-particle operator that is
invariant under rotations around the $z$-axis,
we can restrict ourselves to considering Slater determinants of
angular momentum eigenfunctions, which
leads to
\beqa\nonumber
\Econf^{\rm Q}(N,Z,B)&\geq&\inf_{\Gamma}\Elin[\Gamma]-\widetilde
D(\rddm,\rddm)-R_L\\ \label{i} &=&E^{\rm DDM}(N,Z,B)-R_L.
\eeqa

\begin{rem}[Magnitude of the \lq\lq exchange term\rq\rq]
One may ask for the optimal magnitude of $R_L$ such that a bound
of the form (\ref{i}) is valid. $R_L$ is an upper bound on the
difference between $\Econf^{\rm Q}$ and $\Eddm$, which is given by
the exchange energy
\beq
I_\Psi=D(\rho_\Psi,\rho_\Psi)-\langle\Psi|\sum_{i<j}|\x_i-\x_j|^{-1}\Psi\rangle,
\eeq
where $\Psi$ is an (approximate) ground state wave function of
$H$. The exchange energy is roughly $N$ times the self energy of
the charge distribution of one particle. This distribution has the
shape of a cylinder with diameter $R\sim B^{-1/2}$ and length
$L\sim \max\{Z^{-2/5} B^{-1/5},Z^{-1}\ln(B/Z^3)^{-1}\}$, if we
assume $N\sim Z$ (see also \cite{LSY95} for heuristic arguments).
Note that $L\gg R$ is equivalent to $B\gg Z^{4/3}$. In this case,
we expect an exchange energy of the order
\beqa\nonumber
Z L^{-1} \ln(L/R)&\sim& Z^{7/5} B^{1/5} \ln(B/Z^{4/3})\quad {\rm
for}\quad
Z^{4/3}\ll B \leq Z^3\\ \label{68}
Z L^{-1} \ln(L/R)&\sim& Z^2 \ln(B/Z^3) \ln(B/Z^2) \quad {\rm for}\quad B\gg Z^3.
\eeqa
Hence the exchange energy should be of order $Z^{1/2} (\Eddm)^{1/2}\times$
some factor logarithmic in $B$. We conjecture that (at least for appropriate
$\Psi$ in the lowest Landau band)
\beq
I_\Psi\leq {\rm const.}\frac 1B \int \rho_\Psi^2 \times\,
\mbox{some factor logarithmic in $B$},
\eeq
which is precisely of the correct order (\ref{68}).
This is in accordance with results on the homogeneous electron gas in a magnetic
field \cite{DG71,FGPY92}.
\end{rem}

\section*{Acknowledgement}
The authors would like to thank Bernhard Baumgartner and Jakob Yngvason
for proofreading and valuable comments.

\end{document}